\documentstyle[prl,aps]{revtex}
\begin{document}
\title{Stabilizing an Attractive Bose-Einstein Condensate by Driving A
Surface Collective Mode}
\author{Arjendu K. Pattanayak,
Arnaldo Gammal\thanks{Present address: 
Instituto de F\'{\i}sica Te\'{o}rica, Universidade Estadual 
Paulista, 01405-900 S\~ao Paulo, Brazil}
Charles A. Sackett\thanks{Present address: 
Time and Frequency Division, National Institute of Standards and Technology,
Boulder, CO 80303} and Randall G. Hulet}
\address{Department of Physics and Astronomy and Rice Quantum Institute, 
Rice University, Houston, TX 77251-1892}
\date{\today}
\maketitle
\begin{abstract}
Bose-Einstein condensates of $^7$Li have been limited in number due 
to attractive interatomic interactions. Beyond this number, the 
condensate undergoes collective collapse. We study theoretically the 
effect of driving low-lying collective modes of the condensate by 
a weak asymmetric sinusoidally time-dependent field. We find that 
driving the radial breathing mode further destabilizes the condensate, 
while excitation of the quadrupolar surface mode causes the condensate 
to become more stable by imparting quasi-angular momentum to it. We 
show that a significantly larger number of atoms may occupy the 
condensate, which can then be sustained almost indefinitely. All effects 
are predicted to be clearly visible in experiments and efforts are 
under way for their experimental realization.
\end{abstract}
Bose-Einstein condensation (BEC) in cold and dilute atomic gases\cite{BEC}, 
provide a new domain for studying nonlinear many-body quantum 
systems\cite{RMP}. 
Consider the condensate wavefunction in the mean-field picture: 
First, the zero-point kinetic energy of the atoms tends to increase the 
size of the condensate wavefunction. 
Second, this outward pressure is balanced by the effect of the confining 
harmonic trap potential $V(\vec{r})$. Finally, the nonlinearity enters 
through the interatomic interactions scaling as $aN_0|\psi|^2$, 
where $a$ is the {\em s}-wave scattering length for the gas and $N_0$ 
is the number of atoms participating in the condensate. The condensate is then 
well-characterized by the product wavefunction $\phi = \sqrt{N_0}\psi$ 
where the Gross-Pitaevskii (GP) equation\cite{RMP} 
\begin{equation}
\label{eq:GP}
i\hbar {\partial \over \partial t} \psi( \vec{r},t) =
\Big[ -\frac{\hbar^2}{2m} \nabla^2 
+ V(\vec{r}) 
+ 
\frac{4\pi \hbar^2 aN_0}{m} |\psi (\vec{r})|^2  
\Big] \psi (\vec{r},t) 
\end{equation}
governs the single-particle wave-function $\psi$ of this weakly-interacting 
Bose condensate at zero temperature.

Among the systems in which BEC has been observed, $^7$Li is unique in 
that it has a negative value for $a$; {\em i.e.}, the effective 
interaction between atoms is attractive. Such attractive interactions 
were initially argued to prevent BEC\cite{Nogo}, but it is now established 
that a static metastable condensate exists as long as the number of 
participating atoms $N_0$ is less than a maximum $N_{\rm s}$\cite{Nmax}. 
This is understood\cite{Collapse-t} to unfold as follows: 
As the $^7$Li Bose gas is cooled below the critical temperature for BEC, 
the number of atoms in the condensate increases steadily with a
kinetically determined `fill rate'. As $N_0$ increases the nonlinear 
attractive interactions grow and the condensate becomes more localized. 
At $N_0\simeq N_{\rm s}$, the condensate undergoes collective collapse
and there is a rapid increase in condensate density. The rates for 
inelastic collision processes, including two-body dipolar collisions and 
three-body recombination, are density dependent and increase rapidly
during collapse. The atoms participating in these processes acquire
large kinetic energies and are ejected from the condensate. 
Although the quantitative details of this collapse are not yet fully 
understood, experimental evidence for this process has been
obtained\cite{Collapse-e}.

In this Letter, we analyze theoretically the effect of weakly perturbing 
the harmonic trapping potential for an attractive condensate. 
We have considered general potentials $V\left(\vec{r}\right)=
{1\over 2}m \sum_{i=1}^{3}\omega_i^2 x_i^2$
where $\vec{r} \equiv (x_1,x_2,x_3)\equiv(x,y,z)$ and the parameters
$\omega_i$ characterize the trapping potential along the three axes. 
For the case presented in detail below, $\omega_i^2 = 
\omega_0^2[1 + \alpha_i\cos(\omega t)]$ where $\omega$ is the frequency
of the forcing and the $\alpha_i$ are its amplitudes. This corresponds 
to a spherically symmetric trap driven asymmetrically by a weak, 
sinusoidally time-dependent field. We employ a Gaussian time-dependent 
variational principle approximation\cite{Collapse-t} (GVA) analysis of 
the GP equation to solve for the condensate dynamics. The Gaussian ansatz 
is motivated by the shape of the ground-state for non-interacting 
particles in a harmonic trap and constrains the condensate wavefunction 
to be of the form $\psi(\vec{r},t)= \psi_x\psi_y\psi_z$
where $\psi_i =\big [2\pi a_i(t)\big]^{-\frac{1}{4}}\exp
\left\{-x_i^2(\frac{1}{4a_i(t)} +ib_i(t))\right\}.$
This ansatz when used in Eq.~(\ref{eq:GP}) yields that the condensate dynamics 
correspond\cite{Eigenvals-1} to a classical effective Hamiltonian.
With the transformation\cite{akp} $a_i(t) = \rho_i^2(t)$ and 
$b_i(t) = \frac{\Pi_i(t)}{2\rho_i(t)}$ this Hamiltonian has the transparent 
form $H_{var} = \sum_i \big [ \frac{\Pi_i^2}{2m} + \frac{\hbar^2}{8m\rho_i^2}
+ \frac{m\omega_i^2\rho_i^2}{2} \big ] + \frac{\hbar^2 a N_0}
{4 \sqrt{\pi}}\frac{1}{\rho_1\rho_2\rho_3}.$
The parameters $\rho_i$ and $\Pi_i$ are given by 
$\langle \hat x_i^2\rangle = \rho_i^2$ and 
$\langle \hat x_i\hat p_i + \hat p_i\hat x_i\rangle = 2\rho_i\Pi_i$.
Note that $\langle \hat x_i\rangle =0 = \langle \hat p_i \rangle$ for this 
wavefunction. The centroid of the condensate thus sits at the trap minimum
and all the dynamics is restricted to changes in the condensate width 
$\rho_i$ in the three directions. That is, the dynamics of a quasi-particle 
with canonically conjugate position and momentum 
variables $\rho_i$ and $\Pi_i$ characterizes the condensate. 
A further scaling by the length of the trap 
$\ell_0=\sqrt{\hbar /(m\omega_0)}$, the time-scale $\omega_0^{-1}$, 
and energy $\hbar \omega_0$ yields the dimensionless Hamiltonian
\begin{equation}
\label{Eq:Hvar}
H_{var} 
= \frac{1}{4}\sum_i \big [ \frac{\Pi_i^2}{2} + \frac{1}{\rho_i^2} +
(1 + \alpha_i\cos(\omega t)) {\rho_i^2} \big ] 
- \frac{3\beta}{4\rho_1\rho_2\rho_3}
\end{equation}
where $\beta = \frac{4}{3\sqrt{2\pi}} \frac{N_0 |a|}{\ell_0}$.
The kinetics of the condensate, including the fill rate and dissipative 
losses, affect the dynamics through the time-dependence of $N_0$. 
We set the fill rate $G_0$ to a constant\cite{fillrate}, which is a good 
approximation to the results from the Boltzmann equation\cite{Collapse-t} 
over the time-scales considered. Further, dipolar relaxation scales as 
$\phi^2$, while molecular recombination is a three-body 
process scaling as $\phi^3$. In particular, with the Gaussian 
approximation for the wavefunction, we can write 
\begin{equation} \label{Eq:kinetics}
\dot{N}_{0} = G_0 
-\frac{N_0^2}{\pi^{3/2}
\rho_1\rho_2\rho_3} 
\left( \frac{G_2}{2\sqrt{2}} + 
	\frac{N_0 G_3}{6 \sqrt{3} \pi^{3/2} 
\rho_1\rho_2\rho_3} 
\right)
\end{equation}
where $G_1$ and $G_2$ are the appropriate rate constants\cite{gerton}. 
The Hamiltonian Eq.~(\ref{Eq:Hvar}) and the kinetics Eq.~(\ref{Eq:kinetics}) 
define the condensate dynamics completely. 

To understand these dynamics consider the fixed points given by 
${\partial H_{var}}/{\partial \Pi_i}=0={\partial H_{var}}/{\partial \rho_i}$.
This yields a solution with $\Pi_i =0$ and non-zero $\rho_i$ corresponding
to the width of a metastable condensate. This solution 
$\rho_\beta = |\vec{\rho}_\beta|$ is strongly affected by the attractive 
hole at the origin ($\vec{\rho} = 0$ or infinite condensate density) 
due to the term $\beta/(\rho_1\rho_2\rho_3)$ in the Hamiltonian. 
The condensate is only stable when the quasi-particle avoids this 
attractive hole. As $N_0$ and correspondingly $\beta$ increase, the 
metastable minimum moves closer to the origin, becomes shallower until 
it becomes an inflection point, and finally vanishes. Correspondingly, the 
width $\rho_\beta$ decreases abruptly when the minimum vanishes, leading to 
collapse via the kinetics. The critical values $\rho_m$ and $\beta_m$ 
for collapse are determined by further imposing the inflection condition 
$\frac{d^2\; H_{var}}{d\; \rho^2_i} = 0$. This gives
$\rho_m = {5^{-1/4}}$ and $\beta_m = \frac{8}{3}{5^{-5/4}}$,
corresponding to $N_{\rm s} \approx 1400$ atoms for the conditions of the
Rice experiment\cite{Collapse-e}. 

The normal modes of the quasi-particle dynamics around this minimum (when 
it exists) are the lowest-lying even collective excitations of the 
condensate\cite{Eigenvals-1,Eigenvals-2,Eigenvals-3}. 
These normal mode frequencies are obtained by computing the matrices 
${\cal V, T}$ with elements ${\cal V}_{ij} 
= \frac{\partial^2 H_{var}}{\partial \rho_i\partial \rho_j},
{\cal T}_{ij} = \frac{\partial^2 H_{var}}{\partial \Pi_i\partial \Pi_j}$ 
and then solving for the roots $\Omega_i$ of the equation 
$\det({\cal V} -\Omega^2{\cal T}) = 0$. 
One solution is $\Omega_1 = \sqrt{5 - \rho_\beta^{-4}}$ which is the 
frequency at which the quasi-particle vibrates radially, corresponding 
to the condensate's `breathing' mode. The other two solutions are 
the degenerate solutions $\Omega_{2,3} = \sqrt{2 + 2\rho_\beta^{-4}}$.
corresponding to vibrational motion perpendicular to the radial direction. 
This can be visualized as being along the surface of a sphere of fixed 
radius -- these are then quadrupolar surface modes
\cite{Eigenvals-1,Eigenvals-2,Eigenvals-3} of the 
condensate. Since $\rho_\beta$ decreases as a function of $N_0$, 
$\Omega_1$ decreases and $\Omega_{2,3}$ increases with increasing $N_0$. 
Also note that $\Omega_{1,2,3} = \Omega_0 = 2$ for $\rho_\beta = 1$,
{\em i.e.} for $\beta =0$ (non-interacting limit) in these natural
units. 

We now turn to the numerical solution of this system of equations. As 
expected, these lowest-lying even collective modes of the condensate are 
excited by the driving. Driving the lowest such mode, a radial 
breathing mode, further destabilizes the condensate and decreases the
maximum number of atoms below the static maximum $N_{\rm s}$. Remarkably, 
driving the next-highest excitations, which are quadrupolar surface modes, 
causes the condensate to become more stable. That is, for certain 
frequencies, atoms occupy the condensate in numbers significantly greater 
than $N_{\rm s}$ and the condensate is sustained for correspondingly 
longer times. We have studied in particular the frequency-dependent 
value of the maximum number of atoms $N_{\omega}$ that can be sustained in the 
presence of the driving. We find that this frequency response is broad 
and qualitatively robust, with similar features for wide range of tested 
parameters and many alternative configurations for both the confining 
and perturbing fields. An example of these responses is shown in Fig.~(1), 
along with a theoretical estimate $N_{\rm ang}$ as explained below. In 
obtaining it, we use $\alpha_3 = 0.02 = -2\alpha_1 = -2\alpha_2$ which 
represents the effect of a Helmholtz field oriented along the axis of 
an Ioffe-Pritchard magnetic trap. The response is shown as the ratio 
$N_{\rm m} = N_{\omega}/N_{\rm s}$. The broad stabilization effect is 
clear in this figure; further, it can be seen that the curve has some 
interesting structure, including in particular a signature dip 
corresponding to a parametric resonance between the radial and 
quadrupolar modes as explained below.
The stabilization effect is so strong that in some cases it is no longer 
obvious whether and when a collapse occurs. It is possible for $N_0$ to 
grow so large that the loss rate equals the fill rate, even for relatively 
large $\rho$. The values in Fig.~(1) are therefore calculated using a 
`worst case' definition of the collapse, as the $N_0$ value at which 
the inelastic decay terms in Eq.~(\ref{Eq:kinetics}) exceed the growth 
rate $G_0$. Although it is unclear whether the physics of the true 
dynamical situation in these cases is fully captured by the simple 
kinetics of Eq.~(\ref{Eq:kinetics}), the stabilization is nonetheless 
remarkable.

To understand the detailed structure of Fig.~(1), consider the case of 
driving the condensate with $\omega < \Omega_0$. Note that the dynamics
of the quasi-particle are always restricted to the $\rho_1 =\rho_2$ plane 
by our choice of driving. At an arbitrary value of $\omega$, the driving 
is not initially resonant with either of the excitations and there is a 
negligible response from the condensate. As $N_0$ increases with time, 
however, $\Omega_1(\beta)$ will ultimately equal any $\omega < \Omega_0$. 
The breathing mode is excited as a result and the quasi-particle 
oscillates along the radial direction. As this oscillation increases in 
amplitude, the quasi-particle is driven into the attractive hole and 
the condensate collapses. The net effect is to decrease $N_m(\omega)$,
with the impact being greatest for $\omega$ slightly less than $\Omega_0$. 
The more interesting case is for $\omega > \Omega_0$ where again, the 
driving is not initially resonant. As $N_0$ increases, however, the 
quadrupolar surface modes come into resonance. The quasi-particle then
oscillates with increasingly greater amplitude in a direction perpendicular 
to the radial direction, still in the $\rho_1 =\rho_2$ plane. An angular 
momentum vector $\vec{l}_{\hat{q}}(t) = \vec{\rho} \times \vec{\Pi}$ 
pointing in the $\hat{q}$ direction can be associated with this motion, 
where $\hat{q} \equiv \frac{1}{\sqrt{2}}(1,-1,0)$ is the normal to 
the $\rho_1 =\rho_2$ plane. This angular momentum vector reverses 
orientation\cite{central} during the oscillation, with its magnitude 
going to zero at the turning points of the oscillation. However, the 
mean-square value of this angular momentum increases as the surface 
mode acquires increasing energy. By virtue of the energy in this 
oscillation, the quasi-particle avoids the chasm of the attractive 
`hole', and the condensate is stabilized, avoiding collapse even when 
the attractive interactions are significant. The above implies that the 
stabilizing effect may be quantitatively estimated by adding to the 
kinetic energy pressure terms an angular momentum term 
$\frac{\bar{l}^2}{2m\rho^2}$ where $\bar{l}^2$ is the time-average of 
$l^2_{\hat{q}}(t)$. This 
yields an estimated increase in the maximum number of atoms as 
$N_{\rm ang}\equiv \frac{N_{\rm m}(\bar{l}\neq 0)}{N_{\rm m}(l=0)} = 
(1 + \frac{4\bar{l}^2}{3})^{5/4}$; this estimate is plotted in Fig.~(1) 
with $\bar{l^2}$ computed directly from the dynamics. It is clear from 
comparing the curves that this simple angular momentum argument captures 
the essential features of the stabilization. In particular, the region 
of discrepancy is precisely where the naive kinetics and collapse 
criterion render suspect the value of $N_{\omega}$ shown, as argued above. 

This resonant driving is easier to sustain for frequencies close to 
$\Omega_0$ since $\Omega_{2,3}$ has a weaker dependence on $N_0$ in that 
regime. As $N_0$ increases, the condensate ultimately falls out of 
resonance with the driving. Consider for example the condensate 
radius $\rho(t) = |\vec{\rho}(t)|$: It is initially unaffected 
by the driving and decreases slowly as in the static case; this is 
followed by a resonantly driven increase that ultimately saturates. 
An example of this mechanism is shown in Fig.~(2a) for $\omega = 2.3$ 
(corresponding to a driving frequency of $\approx 308Hz$ for the Rice 
experiment). In Fig.~(2b) we show a numerical solution for the GP 
equation corresponding to the situation in Fig.~(2a); the two curves 
agree qualitatively. We note here that the numerical solution to 
the nonlinear Schr\"odinger equation in the GP form for cylindrical 
symmetry is extremely difficult and a full scan of frequencies is 
computationally prohibitive. However, simulations including filling 
and inelastic decay similar to those in the literature\cite{Kagan} 
have been performed to qualtitatively validate the GVA results. 

As $\omega$ increases, the resonance occurs at increasingly later times 
and for decreasing windows of time, and hence the greatest stabilization 
happens for $\omega$ slightly greater than $\Omega_0$. There is no such 
stabilization for $\omega > \omega_{2} = \sqrt{12} = 3.46$; this critical 
frequency is obtained by substituting the maximum value for the
equilibrium $\rho_m$ into the expression for $\Omega_{2,3}$ -- {\em i.e.}, 
the driving cannot resonate with the condensate modes if the metastable 
minimum does not exist in the first place. This cut-off can be seen clearly 
in Fig.~(1). Finally, we point out the significant dip in the curves at 
$\omega = \omega_{1} = \sqrt{8}$. This frequency corresponds to the 
situation where $\Omega_{2,3}(\beta) = 2\Omega_1(\beta)$ {\em i.e.} for 
$\rho_\beta^{-4}=3$.  In this case, the driving and the surface modes 
parametrically excite the radial mode through the $1:2$ resonance between 
the frequencies.  Thus, the condensate is destabilized even for a 
frequency greater than $\Omega_0$. Similar resonances account for the 
other detailed structure in the frequency response curves. All these 
features should be clearly visible experimentally, since the experimental 
resolution is of the order of $50$ condensate atoms\cite{Collapse-e}, and 
the predicted features exceed this resolution significantly. Moreover, 
the details of the curves are sensitive to the precise models and parameters 
used. Comparison of experimental results with these theoretical predictions 
will therefore help a) define the regime of validity of a given model and 
b) improve our understanding of the details of the rich interplay between 
kinetics and non-linear many-body quantum dynamics in the attractive 
condensate. Efforts are under way for experimental realization of these 
phenomena.

The instability and finite lifetimes of attractive condensates have been 
a constraint in the push to understand these systems. It is expected that 
the significantly increased stability through weak driving will open the 
way to further novel experimental phenomena and interesting theory and 
ultimately a deeper understanding of this nonlinear quantum regime.

We gratefully acknowledge Keiko Petrosky and Hilary Lovett for their
contributions to this research. AKP benefitted from useful discussions 
with Paul Stevenson. Research at Rice was supported by the NSF, ONR, 
the Welch Foundation and NASA. AG aknowledges partial support from the 
Funda\c{c}\~ao de Amparo \`a Pesquisa do Estado de S\~ao Paulo.

\begin{figure}[htbp]
\caption{Stabilization of an attractive condensate by weak driving as 
measured by the change in the maximum number of atoms participating 
in the condensate as a function of the driving frequency. The ratio 
$N_{\rm m}$ between the frequency-dependent maximum number of atoms 
$N_{\omega}$ and the maximum value $N_{\rm s}$ for the static condensate
is shown. The figure is truncated at $N_{\rm m} \approx 4.8$ due to the
excessive ($ t > 100 s$) amount of time taken to reach this value.
An estimated value $N_{\rm ang}$ for $N_m$ obtained from an angular momentum 
analysis is also shown. See text for details.}
\end{figure}

\begin{figure}[htbp]
\caption{(a) Time-dependence of the radially-averaged width of the condensate
$\rho(t)$. There are rapid oscillations, as expected, on the time-scale of 
the driving ($308 Hz$) that cannot be easily resolved on the scale of this
figure. (b) As in (a) except as computed with a numerical solution of
the GP equation with added dissipation. See text for details.}
\end{figure}

\begin{thebibliography}{99}
\bibitem{BEC}
M.~H. Anderson {\it et~al.}, Science {\bf 269},  198  (1995);
C.~C. Bradley, C.~A. Sackett, J.~J. Tollet and R.~G. Hulet,  
Phys. Rev. Lett. {\bf 75},  1687  (1995);
K.~B. Davis {\it et~al.}, {\em ibid} {\bf 75},  3969  (1995).

\bibitem{RMP}
F. Dalfovo, S. Giorgini, L.~P. Pitaevskii and S. Stringari,  
Rev. Mod. Phys. {\bf 71}, 463 (1999) and extensive references therein.

\bibitem{Nogo}
See for example H.~T.~C. Stoof, Phys. Rev. A {\bf 49},  3824  (1994);
N. Bogolubov, J. of Phys. {\bf XI},  23  (1947).

\bibitem{Nmax}
P.~A. Ruprecht, {\em et al.}, Phys. Rev. A {\bf 51},  4704  (1995); 
C.~C. Bradley, C.~A. Sackett and R.G.~Hulet, Phys. Rev. Lett. {\bf 78},  
985 (1997).

\bibitem{Collapse-t}
C.~A. Sackett, H.~T.~C. Stoof and R.~G. Hulet, Phys. Rev. Lett. {\bf 80}, 
2031  (1998).

\bibitem{Collapse-e}
C.~A. Sackett, J.~M. Gerton, M. Welling and R.~G. Hulet, \prl {\bf 82},
876 (1999).

\bibitem{fillrate} The qualitative results have been verified to not
depend on this assumption. The quantitative effects will be considered 
in detail elsewhere.

\bibitem{gerton}J.~M. Gerton, C.~A. Sackett,B.~J. Frew and 
R.~G. Hulet, \pra {\bf 59}, 1514 (1999).

\bibitem{Eigenvals-1}
V.~M. Perez-Garcia {\em et al.}, \prl {\bf 77}, 5320 (1996).
\bibitem{Eigenvals-2}
K. Singh and D. Rokhsar, Phys. Rev. Lett. {\bf 77},  1667  (1996).
\bibitem{Eigenvals-3}
M. Edwards {\em et al.}, \prl {\bf 77}, 1671 (1996). 

\bibitem{akp}A.~K. Pattanayak and W.~C. Schieve, \pre {\bf 50}, 3601 (1994).

\bibitem{Kagan} Yu. Kagan, A.~E. Muryshev and G.~V. Shlyapnikov, 
\prl {\bf 81}, 933 (1998).

\bibitem{central} Note that this is not a central force problem and as
such the angular momentum is not conserved.
\end{thebibliography}
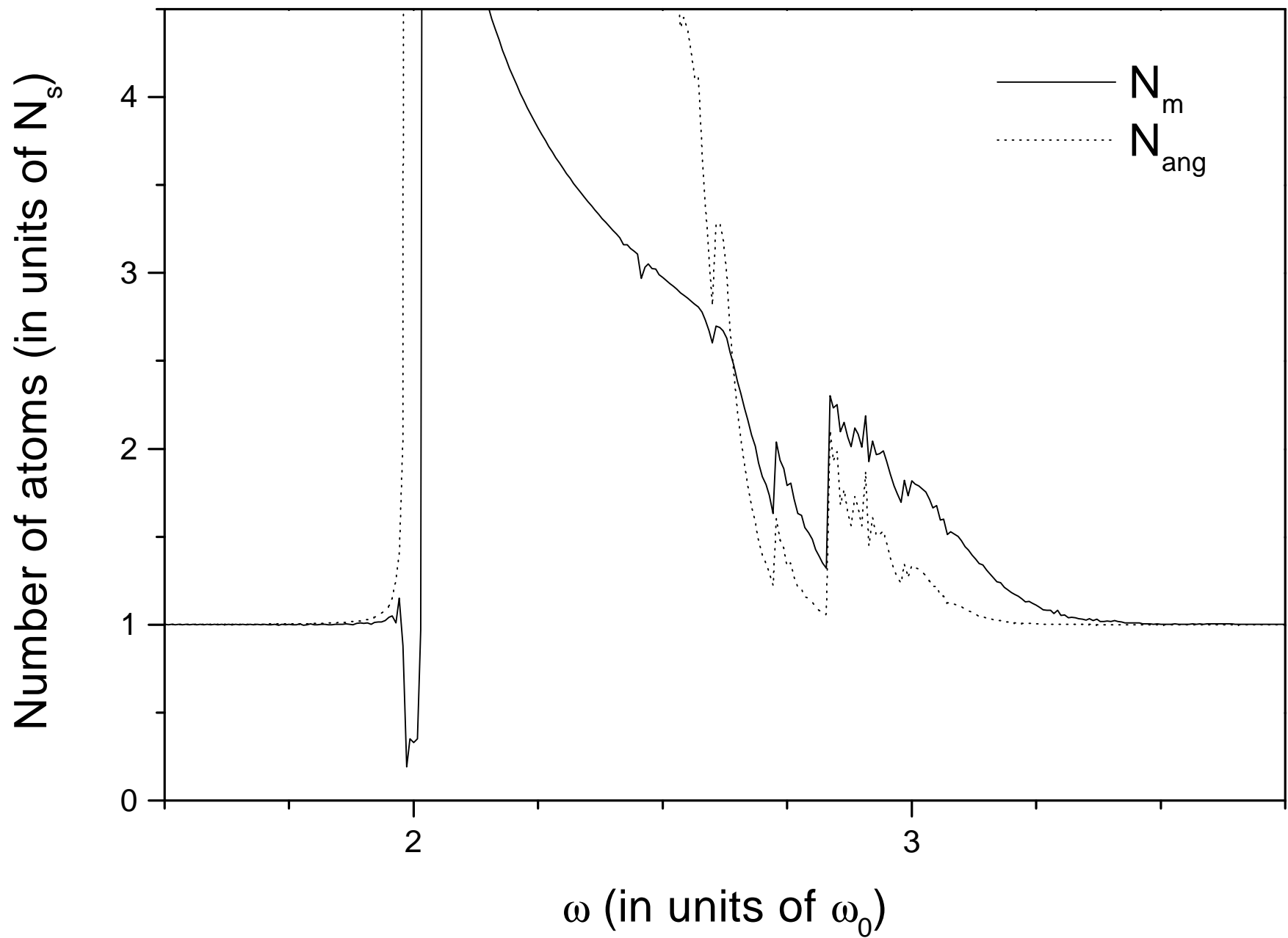
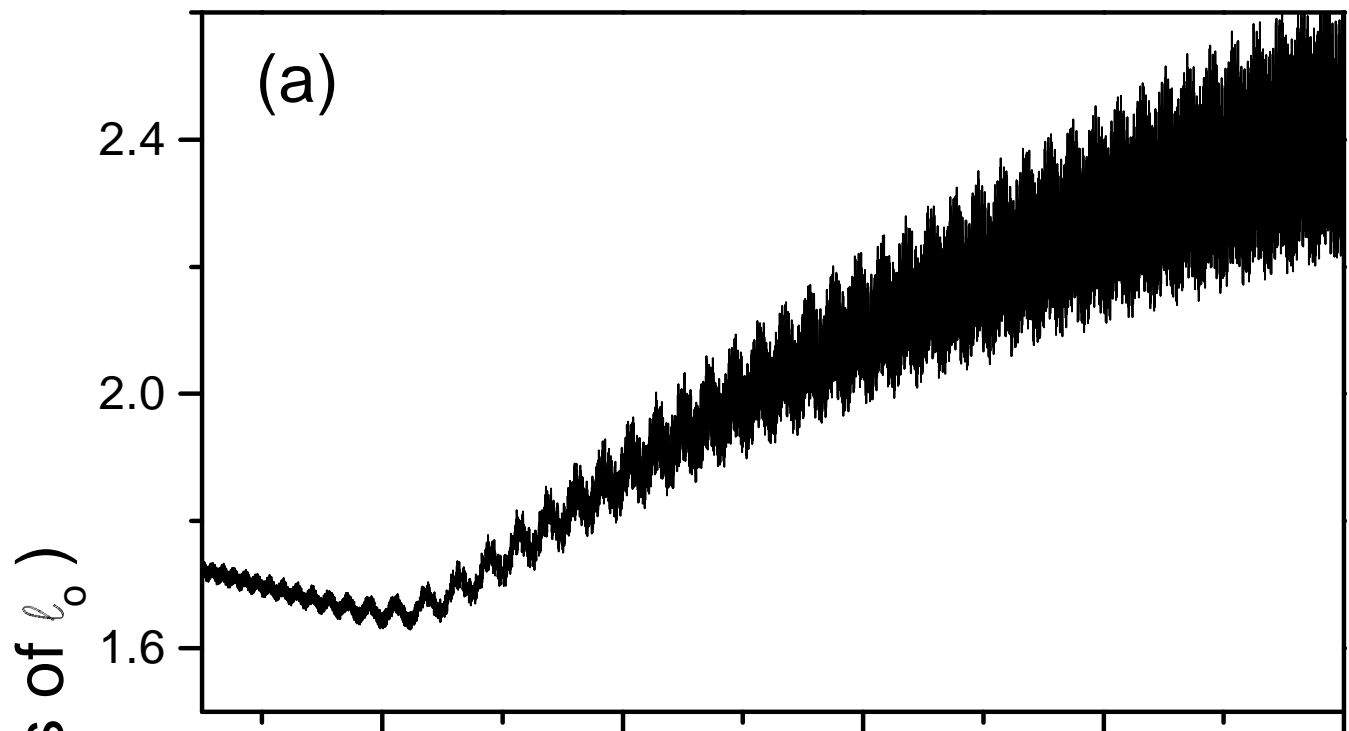
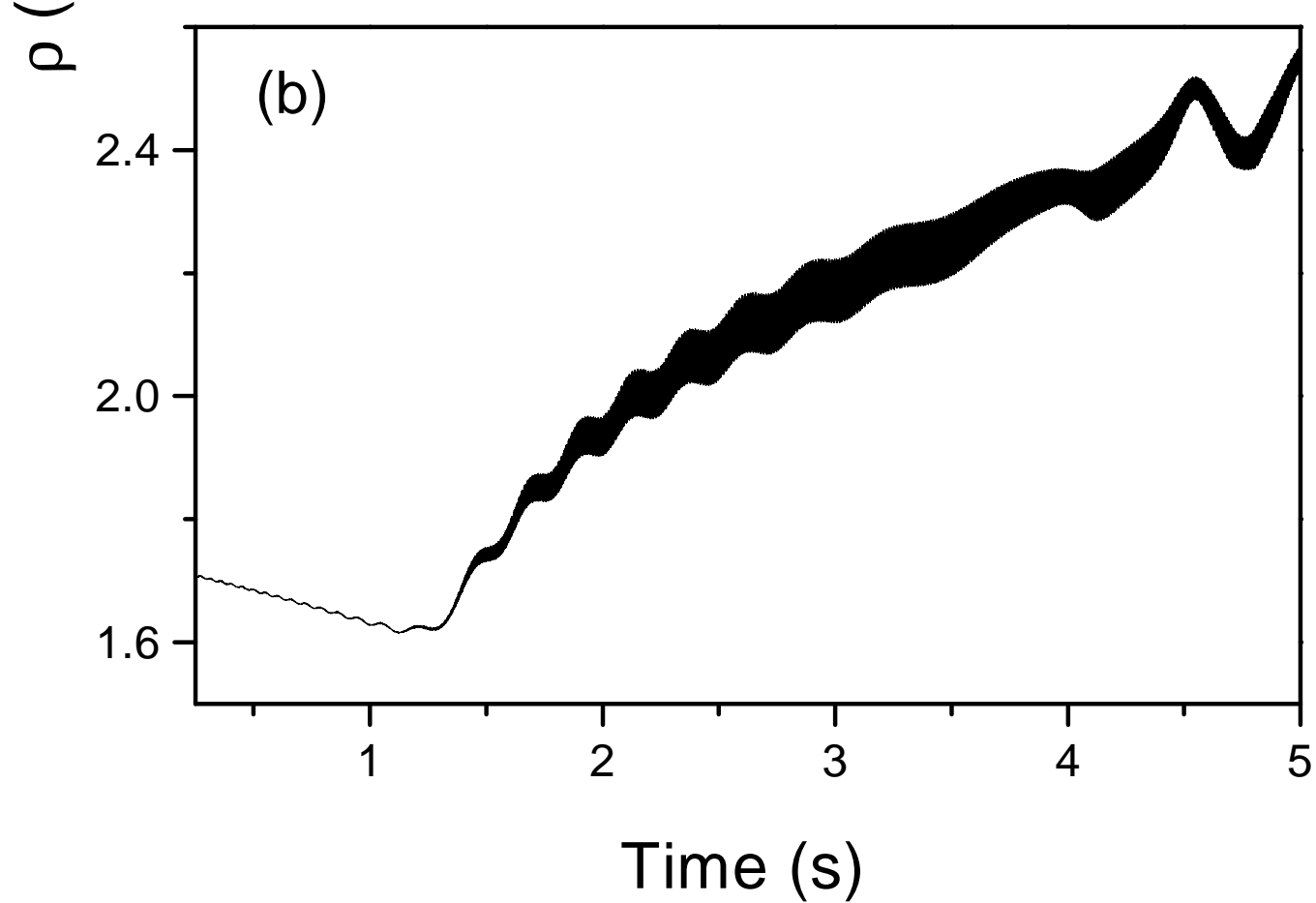
\end{document}